# Monte Carlo simulation of scattered circularly polarized light in biological tissues for detection technique of abnormal tissues using spin-polarized light emitting diodes

Nozomi Nishizawa[1]*, Atsushi Hamada[1], Kazumasa Takahashi[1], Takahiro Kuchimaru[2] and Hiro Munekata[1]

[1] *Laboratory for Future Interdisciplinary Research and Technology, Tokyo Institute of Technology, Yokohama 226-8503, Japan.*

[2] *Center for Molecular Medicine, Jichi Medical University, Tochigi 329-0498, Japan.*

E-mail: nishizawa.n.ab@isl.titech.ac.jp

The circular polarization of light scattered by biological tissues provides valuable information and has been considered as a powerful tool for the diagnosis of tumor tissue. We propose a non-staining, non-invasive and *in-vivo* cancer diagnosis technique using an endoscope equipped with circularly polarized light-emitting diodes (spin-LEDs). We studied the scattering process of the circularly polarized light against cell nuclei in pseudo-healthy and cancerous tissues using the existing Monte Carlo method. The calculation results indicate that the resultant circular polarizations of light scattered in pseudo tissues shows clear difference in a wide range of detection angle, and the sampling depth depends on those detection angles. The structure of the endoscope probe comprising spin-LEDs is designed based on the calculation results, providing structural and depth information regarding biological tissues simultaneously.

## 1. Introduction

The polarization of light is one of the most remarkable phenomena in nature and has led to numerous discoveries and applications in the field of optics, due to its unique spectral and coherence properties[1]. The state of polarization is varied owing to interaction with matter, and has been utilized to characterize and analyze materials through thin film, multilayered film, and transparent media using an ellipsometry technique[2, 3]. In contrast, polarization of light multiply scattered by a turbid medium such as biological tissues provides valuable information about scattering particles[4, 5]. Penetrating  and propagating in a turbid medium, polarized light is multiply scattered and subsequently depolarized. The degree of depolarization significantly depends on the size and shape of the scatterers as well as the number of scattering events associated with the density and distribution of the scatterers in the medium. Because Bickel *et al*. reported that adjacent different structural tissues can be distinguished and time-dependent structural changes can be evaluated by measuring the polarization state of light scattered from suspensions of biological scatterers4), numerous studies observing biological tissues using the polarization of back-scattered light have been performed[6-13]. Backman *et al*., reported that the potentially useful polarization light scattering spectroscopy technique through the demonstration of an in-situ method of probing the structure of living epithelial cells[5]. Pierangelo *et al*., showed microscopic polarimetric imaging with orthogonal state contrast  for uterine cervical cancer and colon cancer[10, 13].

In most of these experiments, linearly polarization was used for the incident state before illumination and for evaluation of polarization after scattering in a medium. However, linear polarization has less persistence than circular polarization in a bio-tissue comprising large cell nuclei compared with the wavelength of incident light (Mie scattering regime)[14, 15]. The observation is caused by the difference in the depolarization process between linear and circular polarizations of light: the depolarization of linear polarized light is the randomization of the polarization plane by superposition of light beams with different propagation directions, whereas randomization of circular polarization is caused by transitions to linear polarization by nearly perpendicular scattering and by superpositions of forward- and back-scattered light with opposite helicity respectively. In the Mie regime, because the scattering is predominantly in the forward direction, the complete randomization of circular polarization requires more scattering events compared with that of linear polarization. In the biological observation using linear polarization, the optical components completely depolarized by multiply scattering in the deep region can be ignored, while the components of light scattered fewer times near the surface can be mostly used. Though these approaches are useful for surface observation, they are limited to the range of their practical applications.

In contrast, Kunnen *et al*., reported that the polarization state of scattered light against incident circular polarization can discriminate between ablated cancerous and non-cancerous tissue in vitro[16). They suggested that changes in nuclei size in cells can be characterized by this technique, which would provide a non-invasive diagnostic approach for early disease detection. To apply this technique to a non-invasive and *in-vivo* observation, a straight optical path into a living body such as a rigid endoscope is required. The transmissions of circular polarized light through a winding optical fiber caused unpredictable depolarization. Qi *et al*., demonstrated a rigid endoscope comprising a stainless-steel sheath, motorized rotation stage with wavelength retarders, polarization analyzer, and a CCD image sensor[17). The application to a flexible endoscope with higher versatility requires monolithic and compact circularly polarized light sources and detectors.

We studied lateral-type spin-polarizing light-emitting diodes (spin-LEDs), which can emit circularly polarized light from the side facets. A spin-LED device comprising a ferromagnetic electrode, an ultra-thin oxide layer, and a GaAs-based double-heterostructure. In this device, spin-polarized electrons are injected from the ferromagnetic electrode through an oxide tunneling layer into a semiconductor structure, and radiatively recombined with holes in an active layer. Spin angular momentum can be transferred to light in the radiative recombination process, resulting in emission of light with angular momentum, that is, circularly polarized light. Lateral-type spin-LEDs require no external magnetic field, in which remnant magnetization state is available for circularly polarized light emission. After working for many years, we finally achieved the electroluminescence with almost pure circular polarization at room temperature without applying an external magnetic field[18). A function to control the helicity of circularly polarized light electrically also has been demonstrated on a lateral-type spin-LED device with a pair of ferromagnetic electrodes with anti-parallel magnetization configuration[19, 20). Additionally, circularly polarized light detection can be realized by applying the opposite process: a circularly polarized light irradiated to a facet of a spin-LED excites spin-polarized electrons in the semiconductor structure and generates a photovoltage according to the parallel/anti-parallel alignment of spins of carriers and magnetization of electrodes[21-24]. These spin-LED devices are called spin-polarized photo-diodes (Spin-PDs) in accordance with its role. Spin-LEDs and Spin-PDs are collectively called Spin-photonic devices.

Using the advantages and functionality of lateral-type spin-LEDs in conjunction with their compactness, we speculate that non-staining, non-invasive, and *in vivo* diagnosis of tumor tissues would be possible if the spin-LEDs could be integrated into the tip of a biopsy probe apparatus such as an endoscope. This instrument would be able to detect a kind of precancerous changes *in vivo* in real time without staining, administering a fluorescent agent, or ablation of tissue. This would lead us to optically specify a slightly lesions that can be detect only by pathological examinations

In view of such prospect, it is necessary to clarify the contribution of optical and structural conditions to the resultant polarization of scattered light in order to prove what kind of precancerous changes the scattered light of circularly polarized light is effective. In this study, we investigate the scattering process of circularly polarized light against a single particle with different radii corresponding to cell nuclei in healthy and cancerous tissues, followed by resultant optical properties for multiple scattering in pseudo biological tissues by modified Monte Carlo (MC) method on the basis of the Mie scattering mechanism. An endoscope which incorporates spin-LEDs is also addressed based on the simulation results.

## 2. Experimental methods

Monte Carlo methods refer to a technique first proposed by Metropolis and Ulam to simulate a physical process using stochastic modeling[25]. In light propavation, the MC technique involves tracing light beams through a medium as they are scattered and absorbed. At each scattering interaction, a new random direction is selected based on the single-scattering phase function at that location. MC techniques have also been adapted to model polarization by tracking changes in the polarization of light as it is scattered. Several MC programs with a different method to track the reference plane for representing the light polarization have been developed[26, 27]. In this study, we adapted the polarized light MC algorithm developed by Ramella-Raman *et al*. to investigate the intensity, the polarization, and the passage distribution of scattered light[28]. This MC algorithm is called "meridian plane MC" because the polarization can be described by the Stokes vector S with respect to the meridian planes that is determined by the light propagation direction and a particular axis.

The polarization state of light is expressed by Stokes vector $S$, $S = (S_0, S_1, S_2, S_3)^T$, where $S_0$, $S_1$, $S_2$, and $S_3$ are the Stokes polarization parameters[29]. The first Stokes parameter $S_0$ describes the total intensity of the optical beam; the second parameter $S_1$ describes the preponderance of linearly horizontal polarized light over linearly vertical polarized light; the third parameter $S_2$ describes the preponderance of linear $+45°$ polarized light over linear $-45°$ polarized light; $S_3$ describes the preponderance of right circularly polarized light over left circularly polarized light. The degree of circular polarization (DOCP) is defined by the equation $DOCP = S_3/S_0$.

In the meridian plane method for single scattering, the scattering angle and the azimuth angle are chosen by the rejection method[30] that generates random variables with a distribution in the Mie scattering process. The propagation direction and the Stokes vector after scattering can then be calculated. These calculation steps for the single scattering process are introduced into the standard MC programs, yielding the polarized MC program. The various forms of carcinoma tissues exhibit some common morphological changes. The cell nuclei become enlarged, denser, darken, and exhibit more pleomorphism. Among these changes, it is expected that the nuclear enlargement contributes the most to the polarization of light scattered in biological tissues. In this study, we carried out the polarized MC simulations for non-cancerous and cancerous pseudo biological tissues in aqueous medium with dispersing particles with a diameter a, a = 6 μm and 11 μm, respectively. These particle sizes correspond to typical nuclear sizes in normal and cancerous cells[5, 16]. The wavelength of incident circularly polarized light was fixed to be 900 nm which is the emitting wavelength of lateral-type spin-LED devices. Fortunately, this wavelength is included in biological optical windows between 800 and 1300 nm, where optical absorption and scattering induced by tissues can be maximally reduced. The optical parameters are listed in Table 1. Various types of cancer have different pathological characteristics accompanied by abnormal increase of organelles or morphological changes. These changes are ignored for simplifications in this study. However, the influence of these changes on the results should be recalled and verified in case of the study for the specific part.

## 3. Results and discussion

Figure 1 shows the scattering angle ($\varphi$) dependence of intensity (left column) and $S_3$ values (right column) of scattered light for a single scattering event by a small particle with a diameter, $a$ = (a) 0.1 nm, (b) 6.0 μm, and (c) 11.0 μm that corresponds to the typical size of small particle in the Rayleigh regime, and of cell nuclei in non-cancerous tissue and cancerous tissue in the Mie regime, respectively. Pure right circularly polarized light beam ($S_3 = +1$) is incident from the direction of $\theta = 180°$ toward a particle located at center, as shown in the inset figure of Fig. 1. Note that the radial coordinate axis in the intensity distribution is linear scale in Fig. 1(a) and logarithmic scale in Fig. 1 (b) and (c). Because the dipole moment excited by incident light on a particle smaller than the wavelength is considered to be a single moment along the electric field vector of incident light, the patterns of radiation from the dipole moment (scattered light) shows symmetric behavior. The intensity of scattered light by a small particle (the left panel of Fig. 1(a)) demonstrates the nearly symmetric distribution for forward and backward scattering, which is almost proportional to $\cos^2 \varphi$. The clear S-shape behavior of $S_3$ values shown in the right panel of Fig. 1(a) also result from the projection of a single dipole moment. In contrast, for a sufficiently larger particle than the wavelength (Mie regime), incident light excites multiple dipole moments with different phases. The interference among a plurality of dipole moments on a sphere results in asymmetric and complex behaviors in scattering angle distributions of intensity and polarization (Fig. 1(b) and (c)). The angular dispersions of scattering intensity demonstrate that forward scattering becomes extremely dominant, which is characteristic of the Mie scattering process. The S-shaped behavior of polarization values disappeared due to the complicated interference, resulting in the complex and fine oscillation behaviors. Considering these oscillation behaviors with the extreme deviation of scattering direction toward forward, the expectation $S_3$ values by a single scattering is close to 1 (initial polarization), and the depolarization by scattering event is attributed to a slight degradation in polarization of the scattered light toward the direction to approximately $\pm 30°$. By increasing the particle diameter, oscillation of $S_3$ behavior becomes finer. In the effective scattering angles $\pm 30°$, extremely slight differences are observed.

These scattering characteristics are introduced into the MC algorithm to simulate the multiple scattering process in biological tissues. The geometry of multiple scattering events is shown in Fig. 2 (a). The circularly polarized light beams are irradiated from a circularly polarized light source (a spin-LED) into pseudo biological tissues with an incident angle $\theta = 1°$. To avoid detection of the reflection light at the surface and the scattered light with a fewer number of scattering events, which does not contain any important information regarding the sample, the circularly polarized light detector is located at a distance of 1 mm from the point of incidence with a detection angle $\varphi$. In the area of a bio-tissue, Fig. 2 (a) shows the distribution of simulated light beam paths under the condition that the detection angle is $\varphi = 25 \pm 5°$, the medium comprises spheres of $a = 6$ μm (a pseudo non-cancerous tissue), and a photon number of incident circularly-polarized light is 100000. Figures 2(b) – (d) show detection-angle $\varphi$ dependence of (b) intensity, (c) degree of circular polarization (DOCP), and (d) sampling depth $L$ for pseudo non-cancerous ($a = 6$ μm) and cancerous ($a = 11.0$ μm) tissues, together with those for turbid tissue with small particles ($a = 0.1$ μm) for comparison (Rayleigh regime). Here, $L$ is defined as the maximum depth at which 30% of detected light beams reaches; in other words, $L$ is the typical sampling depth. The characteristics of light scattered in a dispersion with small particles are featureless due to symmetric scattering shown in Fig. 1(a), whereas the simulation results for pseudo biological tissues show specific behaviors according to particle sizes. The intensity of scattered light is nearly the same for both pseudo tissues for almost entire range of $\varphi$, and shows a peak at around $\varphi = 45°$. The dominant forward scattering in Mie regime causes the intensity peak at comparatively large angles, $\varphi \sim 45°$, in contrast with the case in Rayleigh regime. The difference in the $P$ values between both pseudo tissues, namely $P \sim 0.2$, is noticeable in the detectable range of $\varphi$ where the sufficient intensity of scattered light can be obtained, $-30 \sim 60°$. The extremely slight difference of polarization values ($\Delta P \approx 0.00 \sim 0.02$) in the effective scattering angles, $\pm 12 \sim 15°$, shown in Fig. 1(b) and (c) are accumulated by several hundred scattering events, causing distinctive difference. The typical number of scattering events are $80 \sim 380$ times. In the tissue model closer to real cases, the scattering with smaller subcellular particles and irregular microstructures should be considered. The circular and linear polarization of light are depolarized by smaller particles with fewer scattering events. However, the depolarized values do not contribute to the polarization values of outgoing light after a large number of scattering events. At $\lambda = 900$ nm, the $P$ values for pseudo non-cancerous tissues are larger than that for pseudo cancerous tissues. The large/small ratio of the $P$ values between both pseudo-tissues depends on the incident angle and the relationship between wavelength and particle size. In the wavelength region shorter than 800 nm, the magnitude relationship of the $P$ values is reversed. These results indicate that tissue identification can be available in a wide range of detection angles without being influenced by changes of S/N ratio due to fluctuations of light intensity. The $L$ (Fig. 2(d)) values increase monotonically with increasing $\varphi$ angles with no difference between pseudo tissues. This behavior indicates that the sampling depth can be tuned by modulating the detection angle, suggesting that a depth affected by cancers can be deduced by scanning the detection angle.

Additionally, we also calculate the dependence of distance $d$ between the point of incidence and detection. The $d$ dependences of intensity, polarization, and sampling depth show almost no difference between both pseudo tissues. Intensity and sampling depth are monotonously decreased as a detection point becomes far from the point of incidence.

We design a structure of the endoscope probe based on these calculation results. Figure 3 shows a schematic cross-section of an optical device assembly chip which consists of one spin-LED for an irradiation and a few spin-PDs for a detection of circularly polarized light. The diameter of an endoscope is assumed to be 9 mm. The parallel distance between the endoscope distal end and the surface of sample is fixed to be approximately 8 mm. Incident circularly polarized light beams are irradiated from a spin-LED source with an incident angle 1° onto the sample surface, propagated through the medium, and subjected to multiple scattering events. Scattered light beams radiated outward from the point 1 mm from the point of incident are reflected against a parabolic mirror whose focus point is coincident with the radiation point and collected by spin-PDs aligned along a vertical line. The variations of detection angle as shown in Fig. 2 (b)–(d) are expected to provide information about the scattering region of different depths with almost the same intensities. Therefore, the polarization states of light detected at spin-PDs located at the near position and the far-off position from the surface of an endoscope distal end, show the state in shallow and deep regions of the bio-tissue, respectively. This structure can investigate the conditions of disease according to the depth simultaneously, yielding the degree of the cancer progression from the surface to deep part of the lesion. Practically, the distance and angle between the probe and surface of lesion do not always possess the ideal spatial relations assumed in Fig. 3. The accurate distance and angle can be measured using several optical distance range meters attached to the tip of the endoscope separately. Therefore, the sampling depths can be estimated from the $d$ dependence together with the dependence of incident and detection angles. To enhance the accuracy of depths, the further calculations for various spatial configurations, including curved and tilted surface of the lesion, are required. Further, meaningful signals of scattered light including polarization information are weak and buried in the noise due to stray light. Spin-LED devices with a pair of magnetic electrodes with anti-parallel configuration can electrically switch the helicity ($\sigma+/\sigma-$) of emitting circularly polarized light[19, 20]. Synchronous detection with the switching signal of a spin-LED, can remove any noise signal, enabling a highly sensitive detection of scattered light.

## 4. Conclusions

We propose the optical cancer diagnosis techniques utilizing circular polarization of light scattered from biological tissues, which is characteristics of the non-staining, non-invasive, and *in-vivo* observation method using spin-LED devices. As basic research for that purpose, we have studied the Mie scattering process of circularly polarized light against cell nuclei in the tissues before and after carcinogenesis using a meridian plane Monte Carlo method. The results of the simulation for the case that an incident angle is almost perpendicular to the surface concluded these three points: the distinct difference in resultant polarization values between normal and cancerous tissues can be obtained independently of scattering angles; the intensity of light scattered from both tissues shows the same tendency; the sampling depth can be fine-tuned by changing scattering angle. The simulation results show the technique we proposed in this paper is effective for the pre-cancerous changes with the enlargement of cell nuclei at least. Based on these simulation results, we designed the endoscope probe consisting of spin-photonic devices. This optical device assembly can be expected to detect the presence of cancer and the degree of cancer progression in the biological tissues by collecting the structural information with different depths simultaneously. It is hoped that the proposed technology and device would pioneer the practical applications of circularly polarized light as well as of semiconductor-based spin-photonic devices.

## Acknowledgments

We acknowledge the financial support from a Grant-in-Aid for Scientific Research (No.17K14104, No.18H03878 and 19H04441) from the Japan Society for Promotion of Science (JSPS), the Cooperative Research Project of Research Center for Biomedical Engineering, Futaba Foundation and Spintronics Research Network of Japan (Spin-RNJ).

## References

1) M. Born and E. Wolf, Principles of Optics (Cambridge University Press, Cambridge, England, UK, 1999).

2) L. D. Barron, Molecular Light Scattering and Optical Activity (Cambridge Press, Cambridge, London, UK, 1982)

3) R. M. A. Azzam and N. M. Bashara, Ellipsometry and Polarized Light (Elsevier, Amsterdam, NL, 1987).

4) W. S. Bickel, J. F. Davidson, D. R. Huffman, and R. Kilkson, Proc. Nat. Acad. Sci USA **73**, 486 (1976).

5) V. Backman, R. Gurjar, K. Badizadegan, I. Itzkan, R. R. Dasari, L. T. Perelman, and M. S. Feld, IEEE J. Se. Top. Quantum Electron. **5**, 1019 (1999).

6) A. H. Hielscher, J. R. Mourant and I. J. Bigio, Appl. Opt. **36**, 125 (1997).

7) L. F. Rojas-Ochoa, D. Lacoste, R. Lenke, P. Schurtenberger and F. Scheffold, J. Opt. Soc. Am. A **21**, 1799 (2004).

8) V. V. Tuchin, L. V. Wang and D. A. Zimnyakov, Optical Polarization in Biomedical Applications (Springer, New York, US, 2006).

9) N. Ghosh and A. Vitlin, J. Biomed. Opt. **16**, 110801 (2011).

10) A. Pierangelo, A. Benali, M. Antonelli, T. Novikova, P. Validire, B. Gayet and A. D. Martino, Opt. Express **19**, 1582 (2011).

11) P. Ghassemi, Pl Lemaillet, T. A. Germer, J. W. Shupp, S. S. Venna, M. E. Boisvert, M. H. Jordan and J. C. Ramella-Roman, J. Biomed. Opt. **17**, 076014 (2012).

12) P. Doradla, K. Alavi, C. Joseph and R. Giles, J. Biomed. Opt. **18**, 090504 (2013).

13) A. Pierangelo, A. Nazac, A. Benali, P. Validire, H. Cohen, T. Novikova, B. H. Ibrahim, S. Manhas, C. Fallet, M.-R. Antonelli and A. -D. Martino, Opt Exp **21**, 14120 (2013).

14) F. C. MacKintosh, J. X. Zhu, D. J. Pine and D. A. Weitz, Phys. Rev. B **40**, 9342 (1989).

15) D. Bicout, C. Brosseau, A. S. Martinez and J. M. Schmitt, Phys. Rev. E **49**, 1767 (1994).

16) B. Kunnen, C. Macdonald, A. Doronin, S. Jacques, M. Eccles and I. Meglinski, J. Biophotonics **8**, 317 (2015).

17) J. Qi and D. S. Elson, Sci. Rep. **6**, 25953 (2016).

18) N. Nishizawa K. Nishibayashi and H. Munekata, Proc. Nat. Acad. Sci USA **114**, 1783 (2017).

19) N. Nishizawa K. Nishibayashi and H. Munekata, Appl. Phys. Lett. **104**, 111102 (2014).

20) N. Nishizawa, M. Aoyama, R. C. Roca, K. Nishibayashi and H. Munekata, Appl. Phys. Exp. **11**, 053003 (2018).

21) H. Ikeda, N. Nishizawa, K. Nishibayashi and H. Munekata, J. Magn. Soc. Jpn. **38**, 151 (2014).

22) R. C. Roca, N. Nishizawa, K. Nishibayashi and H. Munekata, Jpn. J. Appl. Phys. **56**, 04CN05 (2017).

23) R. C. Roca, N. Nishizawa, K. Nishibayashi and H. Munekata, Proc. SPIE **10357**, 103571C (2017).

24) R. C. Roca, N. Nishizawa, K. Nishibayashi and H. Munekata, J. Appl. Phys. **123**, 213903 (2018).

25) N. Metropolis and S. Ulam, J. Am. Stat. Assoc. **44**, 335 (1949).

26) S. Chandrasekhar, Radiative Transfer, (Oxford Clarendon Press, Oxford, England, UK, 1950).

27) S. Bartel and A. H. Hielsher, Appl. Opt. **39**, 1580 (2000).

28) J. C. Ramella-Roma, S. A. Prahl and L. Jacques, Opt. Exp. **13**, 4420 (2005).

29) E. Collett, Field Guide to Polarization (SPIE, Bellingham, WA, 2005).

30) W. H. Press, et al., Numerical Recipes in C, the art of Scientific Computing (Cambridge University Press, Cambridge, England, UK, 1992).

## Figure Captions

**Fig. 1.** Scattering angle ($\varphi$) dependence of intensity (left column) and $S_3$ values (right column) of scattered light for a single scattering event by a small particle with a diameter, *a* = (a) 0.1 μm, (b) 6.0 μm, and (c) 11.0 μm. Pure right circularly polarized light beam ($S_3 = +1$) is incident from the direction of $\theta = 180°$ toward a particle located at center. Note that the radial coordinate axis in the intensity distribution is linear scale in (a) and logarithmic scale in (b) and (c).

**Fig. 2.** (a) Geometry of multiple scattering events in pseudo biological tissues. The detection-angle $\varphi$ dependence of (b) intensity, (c) degree of circular polarization, and (d) sampling depth *L* for pseudo cancerous tissues (a = 11.0 μm: blue plots and lines) and non-cancerous tissues (a = 6 μm : red ones), together with those for turbid tissue with small particles (a = 0.1 μm: black ones) for comparison.

**Fig. 3.** Schematic cross-sectional design of an endoscope probe structure consisting of one spin-LED, a few spin-PDs and a parabolic mirror for the ideal case that the focus of mirror is at the surface of lesion. The scattered light beams from the detection point (focus of the parabolic mirror) with different angle are reflected at the mirror and detected by respective spin-PDs. When the distance between surface of lesion and endoscope distal end is ~ 8 mm, range of detection angle is $0 < \varphi < 30°$.

**Table I**. Optical parameters used in the calculations. The values of absorption and scattering coefficient are obtained experimentally by human bowel tissues.

| Parameters | | Values |
|---|---|---|
| Refractive index of particle | $n_{particle}$ | 1.59 |
| Refractive index of matrix | $n_{matrix}$ | 1.33 |
| | | |
| Absorption coefficient | $\mu_a$ | 0.10 mm$^{-1}$ |
| Scattering coefficient | $\mu_s$ | 6.86 mm$^{-1}$ |
| | | |
| Diameter of cell nucleus | | |
| in normal tissues | | 6.0 μm |
| in cancerous tissues | $a$ | 11.0 μm |

N. Nishizawa *et al.*,

Figure 1

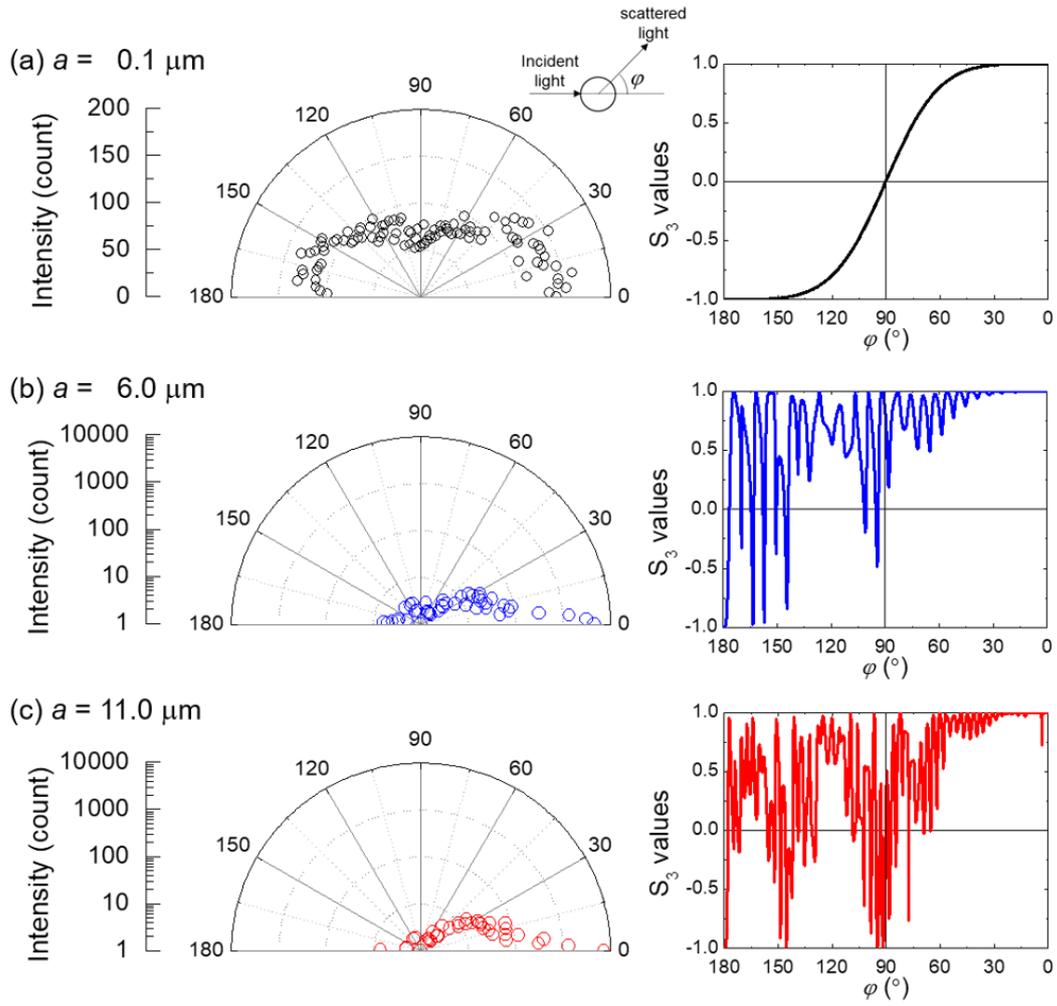

(a) $a =$ 0.1 µm

(b) $a =$ 6.0 µm

(c) $a =$ 11.0 µm

N. Nishizawa *et al.,*

Figure 2

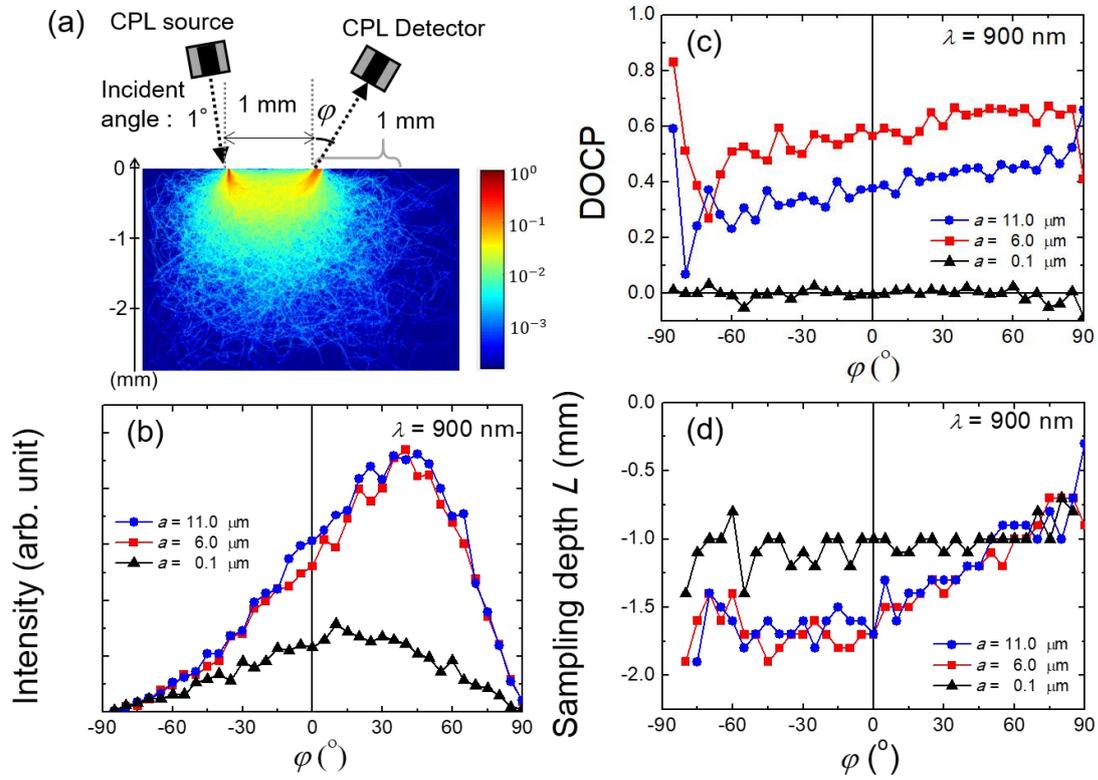

N. Nishizawa *et al.,*

Figure 3

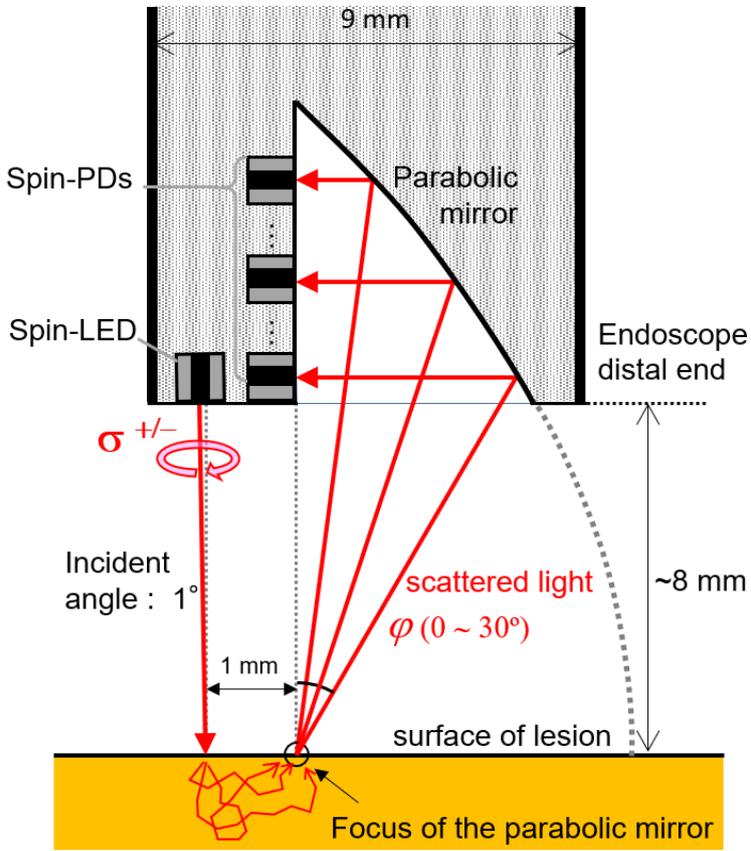

N. Nishizawa *et al.,*